\documentclass[conference]{IEEEtran}
\IEEEoverridecommandlockouts
\usepackage{cite}
\usepackage{amsmath,amssymb,amsfonts}
\usepackage{algorithmic}
\usepackage{graphicx}
\usepackage{textcomp}
\usepackage{xcolor}
\usepackage{float}
\usepackage{multicol}
\usepackage{subcaption}
\usepackage{url}
\usepackage{breakurl}
\usepackage[breaklinks]{hyperref}

\usepackage[section]{placeins}
\def\BibTeX{{\rm B\kern-.05em{\sc i\kern-.025em b}\kern-.08em
    T\kern-.1667em\lower.7ex\hbox{E}\kern-.125emX}}

\begin{document}
%\title{Polyglot Prevention via Proliferation Project
%}
\title{Toward the Detection of Polyglot Files}

\author{\IEEEauthorblockN{Luke Koch\\Bredesen Center\\kochlr@ornl.gov}
\bigskip
\IEEEauthorblockN{Sean Oesch\\Oak Ridge National Laboratory\\oeschts@ornl.gov}
\and
\IEEEauthorblockN{Mary Adkisson\\Tennessee Tech University\\adkissonma@ornl.gov}
\bigskip
\IEEEauthorblockN{Sam Erwin\\Pacific Northwest National Laboratory\\Samantha.erwin@pnnl.gov}
\and
\IEEEauthorblockN{Brian Weber\\Oak Ridge National Laboratory\\weberb@ornl.gov}
\bigskip
\IEEEauthorblockN{Amul Chaulagain\\Oak Ridge National Laboratory\\chaulagaina@ornl.gov}

\IEEEcompsocitemizethanks{\IEEEcompsocthanksitem{ Notice: This manuscript has been authored [or, co-authored] by UT-Battelle, LLC, under contract DE-AC05-00OR22725 with the US Department of Energy (DOE). The US government retains and the publisher, by accepting the article for publication, acknowledges that the US government retains a nonexclusive, paid-up, irrevocable, worldwide license to publish or reproduce the published form of this manuscript, or allow others to do so, for US government purposes. DOE will provide public access to these results of federally sponsored research in accordance with the DOE Public Access Plan (\url{http://energy.gov/downloads/doe-public-access-plan}).}}

}

\maketitle
\begin{abstract}
Standardized file types play a key role in the development and use of computer software. 
However, it is possible to abuse standardized file types by creating a file that is valid in multiple file types. 
The resulting polyglot (\textit{many languages}) file can confound file type identification, allowing elements of the file to evade analysis.
This is especially problematic for malware detection systems that rely on file type identification for feature extraction. 
%File type identification processes that depend on file signatures can be easily evaded thanks to flexibility in the type specifications of certain file types. 
Although work has been done to identify file types using more comprehensive methods than file signatures, accurate identification of polyglot files remains an open problem. Since malware detection systems routinely perform file type-specific feature extraction, polyglot files need to be filtered out prior to ingestion by these systems. Otherwise, malicious content could pass through undetected. To address the problem of polyglot detection we assembled a data set using the \texttt{mitra} tool. We then evaluated the performance of the most commonly used file identification tool, \texttt{file}. Finally, we demonstrated the accuracy, precision, recall and F1 score of a range of machine and deep learning models. Malconv2 and Catboost demonstrated the highest recall on our data set with 95.16\% and 95.45\%, respectively. These models can be incorporated into a malware detector's file processing pipeline to  filter out potentially malicious polyglots before file type-dependent feature extraction takes place. 
\end{abstract}

\section{Introduction}
\subsection{Overview}

A polyglot is a form of steganographic file that can be successfully interpreted in two or more types\cite{corkami_gtfo, saumil_gtfo, ortiz2019hipaa}.
In other words, a JPG+JAR polyglot is one that presents an image when interpreted by an image viewer and executes self-extracting Java code when fed to the Java Runtime Environment. 
Many combinations are possible, including a polyglot that can be interpreted as an image, a video, a PDF, and a video game depending on which program interprets the polyglot\cite{corkami_gtfo}. 
As we demonstrate in our results (section \ref{results}), common utilities for identifying files do not reliably identify polyglots. 
Ergo, malware detection systems that rely on common utilities for file type identification prior to  feature extraction are vulnerable to polyglot files. 
If a detector recognizes only the JPG portion of a JPG+JAR polyglot, then two problems can arise.
\begin{itemize}
    \item For ML/DL-based detecters, feature extraction will only extract fields from the JPG portion of the file
    \item For signature-based detectors, only malware signatures associated with JPG files will be matched\cite{abusing}
\end{itemize}
In either instance, malicious code in the JAR portion passes through without accurate analysis.
In 2019, researchers at Oak Ridge National Laboratory established that multiple commercial off-the-shelf (COTS) malware detection systems failed to detect 100\% of polyglot malware in the data set\cite{bridges2020beyond}. These polyglots were created by Assured Information Security as part of their red team campaign. Motivated by this failure, we compiled a data set of normal and polyglot files, trained multiple classifiers on this data set, and evaluated the recall and F1 score of these classifiers. Our objective is to protect malware detection systems that rely on file type identification for feature extraction or signature sub-selection by developing a model to filter polyglots out of the file processing pipeline. This will allow COTS tools to filter out potentially malicious polyglots while still benefiting from the rich information provided by file type-specific feature extraction\cite{anderson2018ember}. 
\subsection{Motivation}

%notes: rearrange related works and polyglot desc into background section
%move data desc into methodology
%make tighter format for data types for size/flow

%graphs
%compare (all) the scikit models
%pick metrics based on other papers
%at least acc,recall, f1 for all models in one graph
%loss graph for malcon2?
%top 2 models comparison (catboost malconv)
%    compare performance by file type for these two- pe problematic for malconv and zips for catboost

%model selection: word it in more exploratory way,i.e., results indicate....
%benefits/downsides

%need future works: model tuning, gen more diverse data, feature selection improvement, don't give away best idea

%limitations? probably part of future work: not enough file types, speed 

Malware detection systems commonly rely on file type-specific features in order to classify an unknown file as either malicious or benign\cite{anderson2018ember}. For example, the header of a PE file contains a great deal of information about the file, e.g., the read/write/execute flags of all sections, the offset to the import function table, and the address of the entry point. In a PDf file the XREF table, if present, is a useful source of information for determining which portions of the PDF could contain malicious activity. If the XREF is not present or consistent with the file contents, then malicious code can be found by scanning for \textit{action} objects that launch Javascript code\cite{pdf_attacks, pdf_hiding}. In any event, knowing the file type allows the detector to extract far more information than otherwise possible. If the file type is not known, then only generic features like byte entropy, n-grams, and strings can be used for classification\cite{anderson2018ember}.

In some instances, malware detectors will ignore files if the file type has no known malicious capability. If a polyglot file is identified as having only one file type, only that type will be used for feature extraction and further analysis. File sections from the undiscovered type will simply be ignored, allowing potential malicious activity to go undetected or misunderstood\cite{abusing}. One solution is to forego file type-specific feature extraction. However, this greatly reduces the number of available manual features\cite{anderson2018ember,saxe}. Deep learning could be used to perform automatic feature extraction that is effective across all desired file types. However, this would require the complete overhaul of a file type-specific malware detection system. Instead, we propose simply adding a polyglot filter as a pre-processing step for existing malware detection systems. Provided the polyglot filter has sufficiently high recall, this solution would require minimal alteration to existing solutions. Furthermore, existing solutions would continue to benefit from the rich features that can be extracted once a file is known to contain only one file type. 

%In the afore-mentioned trial, ORNL researchers from the Cybersecurity Research Group (CRG) built a virtual range in order to test the efficacy of COTS tools for the U.S. Navy. The range was composed of virtual machines (VMs) connected by a simulated network. The VMs executed scripts that mimicked real-life user behavior, such as sending emails and using word-processing software, according to a Markov model generated by the group. The network played recorded traffic from the laboratory to create realistic background activity and delivered malware to each of the VMs. Each VM had one of four COTS endpoint malware detection systems installed on them to evaluate the effectiveness of these detection systems in a realistic scenario. 

%Per the laboratory's agreement with the producers of each malware detector, we are not at liberty to disclose their identity. We can relate, however, that all of the detectors in question failed to detect 100\% of the polyglot malware that was transmitted over the network. These polyglots were created by Assured Intypeion Security, a cybersecurity research/evaluation company that specializes in mimicking sophisticated advanced persistent threats (APTs). Given the high failure rate across the board, we sought a solution that could be implemented across the industry with minimal alteration to existing systems.

Aside from the afore-mentioned trial at ORNL, additional evidence of the pressing need for polyglot detection is provided by a 2019 attack on DICOM files\cite{dicom-poly,dicom-survey}. DICOM files are commonly used medical imaging files. The attack involves the creation of a PE+DICOM polyglot that functions as expected when loaded into medical imaging software, yet is also capable of execution as a Windows PE file. The DICOM type is intentionally flexible; the authors anticipated the creation of TIFF+DICOM polyglots for non-malicious purposes\cite{dicom-intent}. However, this flexibility lends itself well to more malicious pursuits when the second file contains malicious code.

Due to the wide variety of file types that exist, we felt that a rule-based approach to polyglot detection would require excessive revision and, at best, only account for previously-encountered forms of polyglots. There are also a large number of discrepancies in terms of malformed files that make the rule-based approach difficult\cite{abusing}.

In an attempt to create a solution that learns the general problem and has the possibility of protecting against novel polyglots, we decided to explore machine and deep learning classifiers. This required the creation of a sizable training set that includes a wide variety of existing polyglots. Before we discuss this data set, we first provide a detailed look at the various types of polyglots that are included within it. 

\section{Background}

\subsection{Polyglot Mechanics}

Although there has been some academic research into file type identification methods (discussed in the related works section \ref{related}), polyglot creation has largely been driven by researchers from industry\cite{corkami_gtfo, saumil_gtfo}. Ange Albertini demonstrated multiple methods for creating polyglots at industry presentations and released the  \textit{python}-native \textit{mitra} tool on Github to enable fellow researchers. We utilized this tool to create our polyglot dataset. Albertini classified polyglots based on the method used to combine donor files into the polyglot. Ergo, \textit{mitra} attempts to create stacks, zippers, parasites, and cavities from pairs of donor files.
\subsubsection{Stacks}
A stack is the simplest type of polyglot. The second file is simply appended to the end of the first file. The caveat is that any byte offsets in the second file may need to be adjusted (increased by the length of the first file) in order for the second file to function as expected. Although there is no restriction on the first file, the second file must not strictly enforce the \textit{magic number at offset zero} rule. A \textbf{magic number} is an arbitrary (hence the \textit{magic} moniker) hexadecimal value that uniquely identifies a file type. This number is used for fast file type identification, since a utility need only scan the first few bytes of the file in order to identify it. However, there are many file types that do not enforce this rule. 

PDF readers commonly accept files as valid PDFs if the magic number is anywhere within the first 1024 bytes of the file\cite{juliawolf, corkami_gtfo} despite the requirement listed in the official documentation\cite{pdf_adobe}. Additionally, some types have no requirement at all for a magic number at offset zero. Zip files commonly begin with a magic number. However, this number is part of the local file header for the first file contained within the Zip archive. According to the specifications\cite{zip_pkware}, the Zip file is indexed via a central directory at the end of the file. This central directory has its own magic number for identification. Ergo, Zip files have no requirement that any magic number be the first byte. This makes them an ideal candidate for the second file in a stack polyglot.
\subsubsection{Parasites}

In a parasite polyglot, the second file is added within comment sections---offset by comment markers---of the first file. Many file types allow for comment sections that are not displayed when the file is interpreted. These comments are only visible when the file is opened for editing by a hex editor like \textit{vim}. In order for the second file within the comment sections of the first file to remain functional, it must not have the strict \textit{magic number at offset zero} rule. The byte offsets for both files must be updated. In the case of the first file, the byte offsets must now account for the second file contents that are now hidden within the comment sections. In the case of the second file, the offsets must account for the fact that the second file contents are now scattered in comment sections of the first file. As long as these offsets are updated correctly, both files will continue to function normally. Although this update process is more complex than the updates necessary for a stack polyglot, there are many file types than can combine to create a parasite. 
\subsubsection{Zippers}

Zippers are a more complex version of a parasite. In a zipper, both files are contained within each other's comment sections. This means that the two donor files must use different markers to begin and end their comment sections. This arrangement is rather unusual in practice, so donor files for zippers are much harder to find. In our data set, only DCM files combined with either GIF or PDF files were able to create zipper polyglots.
\subsubsection{Cavities}

Cavities are created when the second file is hidden within null-padded areas of the first file. This arrangement is only possible when the first file is of an executable or ISO type, wherein memory is allocated in chunks. Since these areas are often null-padded to a standard size, executable or ISO files may contain enough null-padded memory to hide a second file in the padded areas. This is very similar to the classic \textit{code caving} technique often used by malware authors to hide malicious code. The distinction between a cavity polyglot and code caving is that the caved material of a cavity polyglot is a complete file that can function correctly when interpreted by an appropriate program. Since the first 16 sectors of an ISO file are left empty, the contents of another file can be placed in the beginning of the ISO file. For executable files, like Windows PE files, the second file would be written into the null-padded trailing areas at the end of sections. No updates will be needed for the first file, but any byte offsets in the second file will need to be updated if the second file does not begin at offset zero of the first file.

\subsection{Related Work}
\label{related}
File type identification has historically been addressed through signatures. The Linux utility \texttt{file} matches files by examining magic bytes in unison with other structural elements. The magic number is an arbitrary hexadecimal value that uniquely identifies a file type. Some file types can be strictly identified simply by their magic number, while others require additional elements to be scanned. JAR files, for instance, are simply Zip files with a MANIFEST.INF file present in the archive. Ergo, \texttt{file} scans for the presence of this file in order to distinguish a JAR file from a Zip file. In either event, \texttt{file} matches each input file to a unique signature. Note, \texttt{file} scans until it detects a match, then halts. There is a \texttt{--keep going} flag, but in our examination it did not allow \texttt{file} to detect polyglots. The results of running \texttt{file} (with this flag active) on our polyglot data set is included in section \ref{results}. The signatures that \texttt{file} utilizes are extremely accurate and efficient in identifying conventional (monoglot) files, which means little research has been done to improve on this winning formula. 

%File fragments, however, may not contain a complete signature, prompting interesting research into file fragment identification. 

%These authors further detail why simple solutions to file parsing or feature extraction do not exist. We summarize them here:

McDaniels and Heydari used byte histograms in concert with three different algorithms to identify file types\cite{mcdaniels03}. Under the first algorithm (\textit{byte frequency analysis} or BFA), they converted files into a fixed length feature vector. All files, regardless of type, can be represented as a sequence of hexadecimal values. Therefore, we can compactly summarize a file's contents by placing the number of times each possible hex value occurs into a 256 (all possible hexadecimal values) character vector where the index corresponds to the byte value. Ergo, the 7th value stored in the vector is the number of times the value 0x7 occurred in the file. This feature is referred to as a byte occurrence vector, unigram, or byte histogram in literature. McDaniels and Heydari then normalized each byte occurrence vector and calculated an average vector for each file type. Each average vector represented a unique file type and was referred to as a \textit{fileprint}. Test files were compared to the \textit{fileprint} to calculate correlation scores. Finally, each test file received its file type label from the most correlated \textit{fileprint}.

%For the \textit{byte frequency cross-correlation} or BFC algorithm, they  calculated cross-correlation scores to measure the average relationships between bytes per file type. This process resulted in a 256x256 matrix that represented each byte co-occurrence in the file. As with the previous method, an average vector for each file type was calculated and files were labeled according to their correlation with  this average vector. 

%Lastly, the \textit{file header/trailer} or FHT method collected $H$ bytes from the beginning of the file and $T$ bytes from the end of the file. For each position in this string of bytes, a one-hot encoding is produced. A one-hot encoding creates a vector 256 characters long (all possible hexadecimal values) where all values are zero except the value that actually occurs. Ergo, if the first byte was 0x7, then the 7th value in the vector would be a 1 while the rest would be 0. Since this encoding is applied at each offset, the resulting matrix is of size $(H+T)$x$256$. Again, an average \textit{fileprint} per file type was calculated using this representation. The algorithms were tested on 120 files representing 30 file types. The results are as follows: 
%\begin{itemize}
%    \item BFA: 27.5\%
%    \item BFC: 45.83\%
%    \item FHT: 95.83\%
%\end{itemize}
%The relatively high accuracy of the header/footer algorithm demonstrates that the location of a byte value plays a significant role in classification accuracy. The byte histogram, on its own, was not very discriminative. 

Li et al. extended Mcdaniel's work by using centroids rather than an average vector as the representation of each file type\cite{li05}. Files were classified based on their Mahalanobis distance from a centroid. The centroids were chosen using the K-means algorithm. %Note, Manhattan distance is  the metric the authors chose when applying K-means. The authors further experimented with two variations. In the first, they used multiple clusters per file type since some file types are assumed to be quite diverse. In the second variation they randomly select 80\% of their training files for use as exemplars. Under this approach, instead of calculating the Mahalanobis distance from an individual file to a cluster, they find the Manhattan distance to the nearest exemplar. This approach had the highest accuracy (99.6\%) by a slim margin. Interestingly, these authors found the highest accuracy by truncating the input down to the first 20 bytes of each file rather than examine the entire file. The lowest accuracy was 82\%, which corresponded to the single cluster with no truncation method. Their data set consisted of 800 files across 8 file types. 

Karresand\cite{karresand}, Veenman\cite{veenman}, Fitzgerald\cite{fitzgerald}, Beebe\cite{beebe2013,beebe2016} all trained machine learning classifiers that used a wide variety of statistical features extracted from files and fragments, which included unigrams, bigrams, bag-of-words, byte rate-of-change, Kolmogorov complexity, and common/longest strings/bytes. Models developed included support vector machines, K-nearest neighbors, and hierarchical clustering. Beebe released an open-source tool, Sceadan, which achieved 73.7\% accuracy on 30 file types and 8 data types by utilizing a linear SVM trained on an input vector of unigrams concatenated with bigrams. Unigrams simply the natural language processing (NLP) term for the byte occurrence count vector or byte histogram. Bigrams are a 256x256 matrix of co-occurrence counts. This data is usually sparse and slow to compute\cite{mittal}.

More recently, deep learning has been used to detect the type of file fragments. File fragments may contain a partial signature or no signature at all, making rule-based approaches useless. Moreover, deep learning has an advantage over other learning algorithms thanks to its automatic feature extraction, which allows deep models to train on raw bytes. Mittal et al. built a deep learning model, FiFTY, to identify file fragments in 2021\cite{mittal}. They trained a one-dimensional convolutional neural network to identify 75 different file types. Although the authors experimented with training neural networks and convolutional neural networks on a vector of statistical features that included Shannon entropy, Kolmogorov complexity, deviation, skewness, kurtosis, and mean values (arithmetic, geometric, harmonic) as features, raw bytes as the only input proved the most accurate. Their model outscored Sceadan (77.5\% acc vs 69.0\% acc) on a data set of 75 file types, which was released along with the FiFTY model. 

%Image classification techniques have also been applied to the fragment classification problem\cite{xu}, although this approach is theoretically unsound. Interpreting a stream of bytes from a binary file as a two-dimensional image introduces a width parameter that assumes an underlying 2D structure which is not present in a binary file.

In terms of cybersecurity, polyglot files are the subject of a small number of research papers\cite{dicom-poly, dicom-survey, abusing}. Jana and Shmatikov detailed a wide variety of obfuscation methods that target discrepancies between how a malware detector parses a file and how the OS interprets the file\cite{abusing}. Some of these discrepancies cause the detector to misidentify the file type of the incoming file, resulting in the wrong subset of malicious signatures being applied to the file. These methods are referred to as Chameleon attacks\cite{abusing}. Werewolf attacks\cite{abusing}, on the other had, exploit discrepancies in how the file is parsed. Jana and Shmatikov describe one type of Werewolf attack as 'ambiguous files conforming to multiple types'\cite{abusing}. These are polyglot files by another name. The authors note that a malicious TAR+ZIP stack-type polyglot was incorrectly classified by 20 out of 36 malware detectors hosted on Virustotal at that time.
%We present methods for detecting polyglots, are which steganographic files that can be correctly interpreted as (at least) two different file types. Some cybersecurity researchers on the industry side have demonstrated that these files can be created and used to bypass detection, but much work needs to be done within the research community to establish reliable methods for detecting these files.

\subsection{Relation to this Work}
Our work builds upon three of the methods presented above. Our data set consists of polyglot files whose basic structure was described by Jana and Shmatikov, albeit with a different nomenclature. Courtesy of Mittal et. al, we train a one-dimensional convolutional neural network on raw bytes. We also train a variety of ML models on the byte count histogram as detailed by McDaniels and Heydari.

\section{Methodology}

\subsection{Polyglot Data Set}
Our dataset, described in Table \ref{Tab:data}, consists of 7 monoglot or normal files types as the negative class and 21 polyglot file types as the positive class with an 80/20 train/test split. We should note that these polyglots do not contain malicious code. Our objective is a polyglot filter, not a malware detector. Additionally, the use of benign polyglots greatly simplifies data sharing. 

\bigskip
 \begin{minipage}{\linewidth}
\begin{table}[H]
\centering
\captionof{table}{Data Set Overview}
\begin{tabular}{|l|l|l|}
\hline
Data Set & Train & Test \\ \hline
Monoglot & 31,199 & 7,799 \\ \hline
Polyglot & 25,210 & 6,303 \\ \hline
\end{tabular}
\label{Tab:data}
\end{table}
\end{minipage}
\bigskip

There are 3120 of each file type in the monoglot training set and 780 of each file type in the test set. The monoglot file types are as follows:

\begin{multicols}{4}
\raggedright
\begin{itemize}
    \item PDF
    \item PNG
    \item GIF
    \item TIFF 
    \item JPG
    \item DCM
    \item JAR
    \item Zip
    \item PE
    \item ISO
\end{itemize}
\end{multicols}
Feeding a subset of the above files to \textit{mitra}, we created 21 different polyglot combinations. There are 1200 of each file type combination in the training set and 300 in test set. 

\begin{multicols}{3}

\begin{itemize}
\raggedright
    \item DCM+GIF
    \item DCM+JAR
    \item DCM+ISO
    \item DCM+PDF
    \item DCM+Zi[
    \item GIF+ISO
    \item GIF+JAR
    \item GIF+Zip
    \item JPG+JAR
    \item JPG+Zip
    \item PE+ISO
    \item PE+JAR
    \item PE+Zip
    \item PNG+ISO
    \item PNG+JAR
    \item PNG+PDF
    \item PNG+Zip
    \item TIFF+ISO
    \item TIFF+JAR
    \item TIFF+PDF
    \item TIFF+Zip
\end{itemize}
\end{multicols}

Since each pair of file types can only be combined in certain ways (stack, zipper, parasite, cavity), the number of each type of polyglot in this data set is not balanced. Our future work includes the production of a data set that is balanced according to polyglot combination method rather than file type. Table \ref{Tab:combos} contains the breakdown of polyglot combination methods in our training and test sets.

\bigskip
\begin{minipage}{\linewidth}

\begin{table}[H]
\centering
\captionof{table}{Data Set Polyglot Combination Method}
\begin{tabular}{|l|l|l|}
\hline
Polyglot Method & Train & Test \\ \hline
Stack           & 10267 & 2574 \\ \hline
Parasite        & 10301 & 2542 \\ \hline
Zipper          & 1795  & 463  \\ \hline
Cavity          & 2847  & 724  \\ \hline
\end{tabular}
\label{Tab:combos}
\end{table}
\end{minipage}
\bigskip

\subsection{Model Selection}
We trained a wide variety of models to identify polyglot files. Traditionally, machine learning models used in cyber security applications rely on manual features that are specific to each file type\cite{souri_survey, hoda_survey}. Since polyglot files can confound file type-specific feature extraction, our models are file type agnostic. In the case of ML models trained through \texttt{scikit-learn}, the input vector consists of a byte histogram. In other words, the feature vector is 256 integers in length where the value at each index corresponds to the number of times that index occurs as a byte value in the file. Since files are stored internally as hexadecimal values (hence 256 possible values for each byte), this feature can be used regardless of file type. The models we tested include random forest, support vector machine, stochastic gradient descent, light GBM, gradient boosting, and CatBoost.

We also trained and tested a deep learning model. Deep learning via a convolutional neural network (CNN) utilizes automatic feature extraction\cite{raff_survey, malconv, malconv2, gilbert}, making it ideally suited for our task.  The deep learning model we chose is designed for binary (two class) classification of binary (compiled) files, namely, Malconv2\cite{malconv2}. This model is a one-dimensional convolutional neural network developed for malware detection whose feature vector consists of the raw bytes. The first Malconv model demonstrated that CNNs can be effective malware classifiers\cite{malconv}. However, it had some issues that prompted the authors to develop a second, more effective model. 

Since Malconv reads in raw bytes as features, the input file must be truncated if it exceeds the maximum capacity of the model. In the second iteration of the model, Malconv2, the authors exploited the sparse nature of temporal max pooling. This allowed them to partition the varying size files into N sections, then collect only the bytes that would actually get updated with nonzero gradient during the backward pass of the training process. This temporal max pooling allows Malconv2 to intake much larger files than the original Malconv. Moreover, temporal max pooling is used in conjunction with a gating mechanism to provide an attention mechanism. This allows Malconv2 to correlate features that are distant from one another in the byte space. This is especially important for polyglots, where two separate headers might be in the same file at a great distance from one another.

We note that one criticism of Malconv is that it pays far more attention to headers than data or code\cite{demetrio2019explaining}. Although this was an undesirable trait in a malware detector, it is useful for a polyglot detector since the header structure is closely tied to the file type. Note, some file types utilizes footers instead of headers or have no distinct header/footer, so learning which parts of a file are most relevant is a challenging problem for a polyglot detector.

Our final method concatenated the byte histogram vector with the mime-type output of the utility \texttt{file}. Since \texttt{file} is very accurate on normal files, we theorized that adding \texttt{file}'s output to the feature vector would lessen the learning load on a model. 

\section{Results \& Discussion}
\label{results}
\begin{figure}[H]
  \centering
  \captionof{figure}{\texttt{file} Performance on Test Data}\par\medskip
   \resizebox{\linewidth}{!}{
  \includegraphics[width=\linewidth]{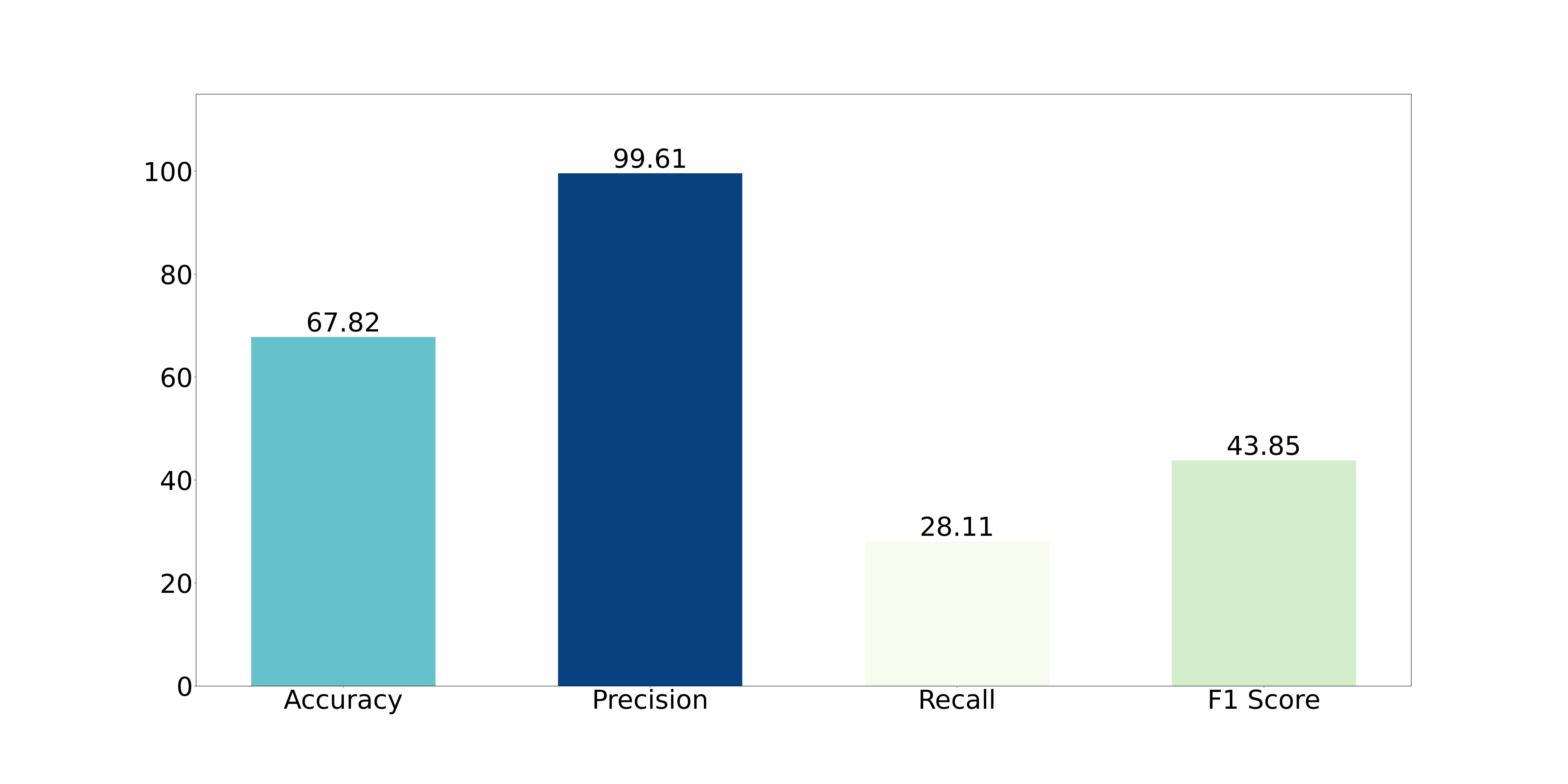}}
  \label{Fig:filebar}
\end{figure}

Firstly, Figure \ref{Fig:filebar} shows that the most common method for identifying file types, \texttt{file},  has poor recall on the test set of monoglot and polyglot files. \texttt{file} does perform extremely well on monoglot (normal) files, as demonstrated by its high precision. Ergo, \texttt{file} is highly effective on expected input, but fails on the most common types of polyglot.

Table \ref{Tab:file} breaks down \texttt{file}'s performance by polyglot type. Although \texttt{file} is effective at detecting zippers and cavities, these two types are the rarest form of polyglot (in our experience) due to the requirements for mutual comment markers and padding bytes, respectively.  We used the \texttt{--keep-going} flag when running \texttt{file}. 

%Figures \ref{file:acc, file:pre, file:rec, file:f1} detail the accuracy, precision, recall, and F1 score of \texttt{file} on each of the four types of polyglots. 
%\begin{figure}[H]
%\label{file:acc}
%  \centering
%  \captionof{figure}{\texttt{file} Accuracy on Test Data Subsets}\par\medskip
%   \resizebox{.8\linewidth}{!}{
%  \includegraphics[width=\linewidth]{images/file_acc.png}}
%\end{figure}

%\begin{figure}[H]
%\label{file:pre}
%  \centering
%  \captionof{figure}{\texttt{file} Precision on Test Data Subsets}\par\medskip
%   \resizebox{.8\linewidth}{!}{
%  \includegraphics[width=\linewidth]{images/file_pre.png}}
%\end{figure}

%\begin{figure}[H]
%\label{file:rec}
%  \centering
%  \captionof{figure}{\texttt{file} Recall on Test Data Subsets}\par\medskip
%   \resizebox{.8\linewidth}{!}{
%  \includegraphics[width=\linewidth]{images/file_rec.png}}
%\end{figure}

%\begin{figure}[H]
%\label{file:f1}
%  \centering
%  \captionof{figure}{\texttt{file} F1 Score on Test Data Subsets}\par\medskip
%   \resizebox{.8\linewidth}{!}{
%  \includegraphics[width=\linewidth]{images/file_f1.png}}
%\end{figure}

\begin{minipage}{\linewidth}
\begin{table}[H]
\centering
\captionof{table}{\texttt{file} Results on Test Data}
\begin{tabular}{|l|l|l|l|l|}
\hline
         & Accuracy & Precision & Recall  & F1 Score \\ \hline
Overall  & 67.82\%  & 99.61\%   & 28.11\% & 43.85\%  \\ \hline
Stack    & 75.75\%  & 90.41\%   & 2.56\%  & 4.99\%   \\ \hline
Parasite & 80.68\%  & 98.75\%   & 21.68\% & 35.55\%  \\ \hline
Zipper   & 99.92\%  & 98.51\%   & 100\%   & 99.25\%  \\ \hline
Cavity   & 99.54\%  & 98.99\%   & 95.58\% & 97.26\%  \\ \hline
\end{tabular}
\label{Tab:file}
\end{table}
\end{minipage}
\bigskip

\begin{figure}[H]
  \centering
  \captionof{figure}{ML/DL Model Accuracy on Test Data}\par\medskip
   \resizebox{\linewidth}{!}{
  \includegraphics[width=\linewidth]{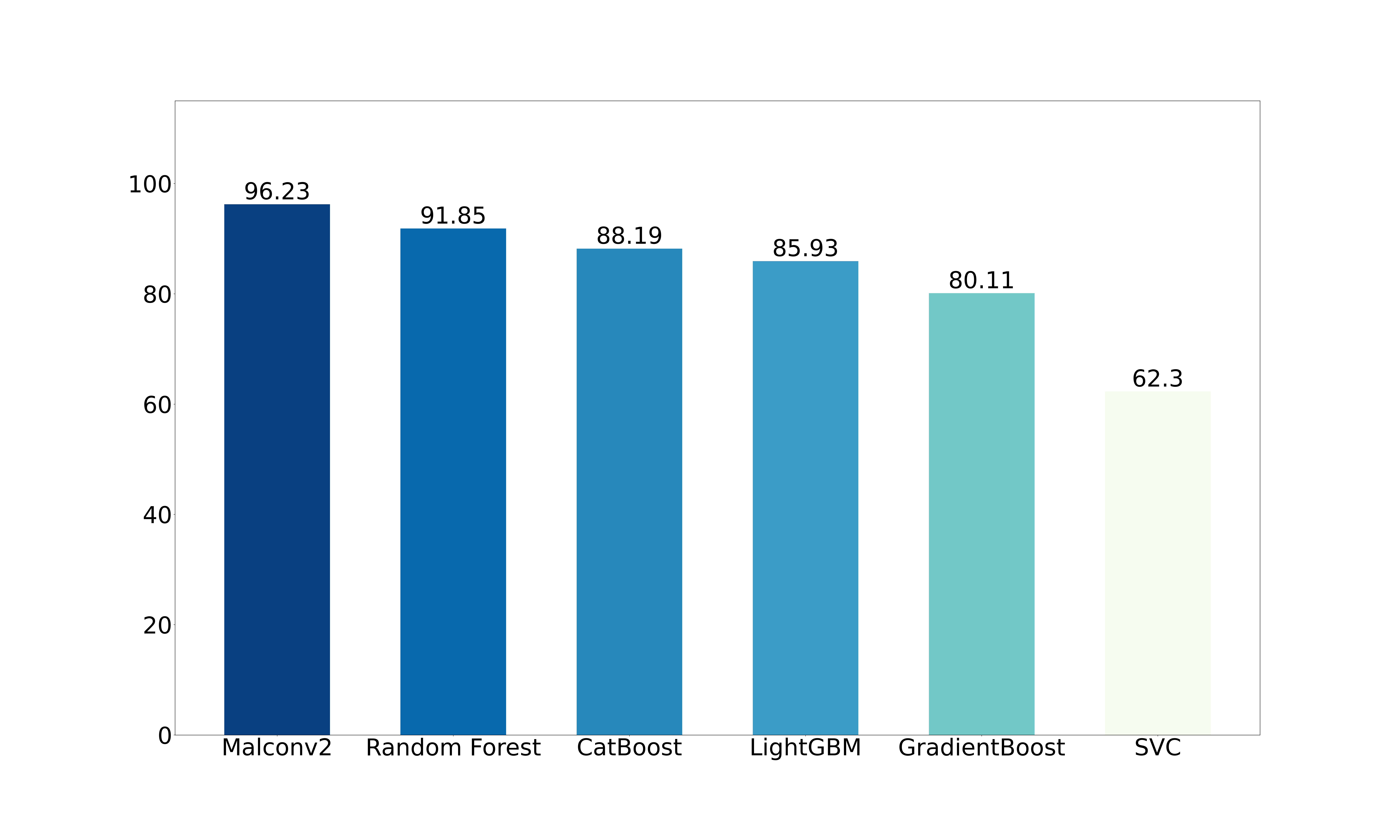}}
  \label{Fig:mlacc}
\end{figure}

\begin{figure}[H]
  \centering
  \captionof{figure}{ML/DL Model Precision on Test Data}\par\medskip
   \resizebox{\linewidth}{!}{
  \includegraphics[width=\linewidth]{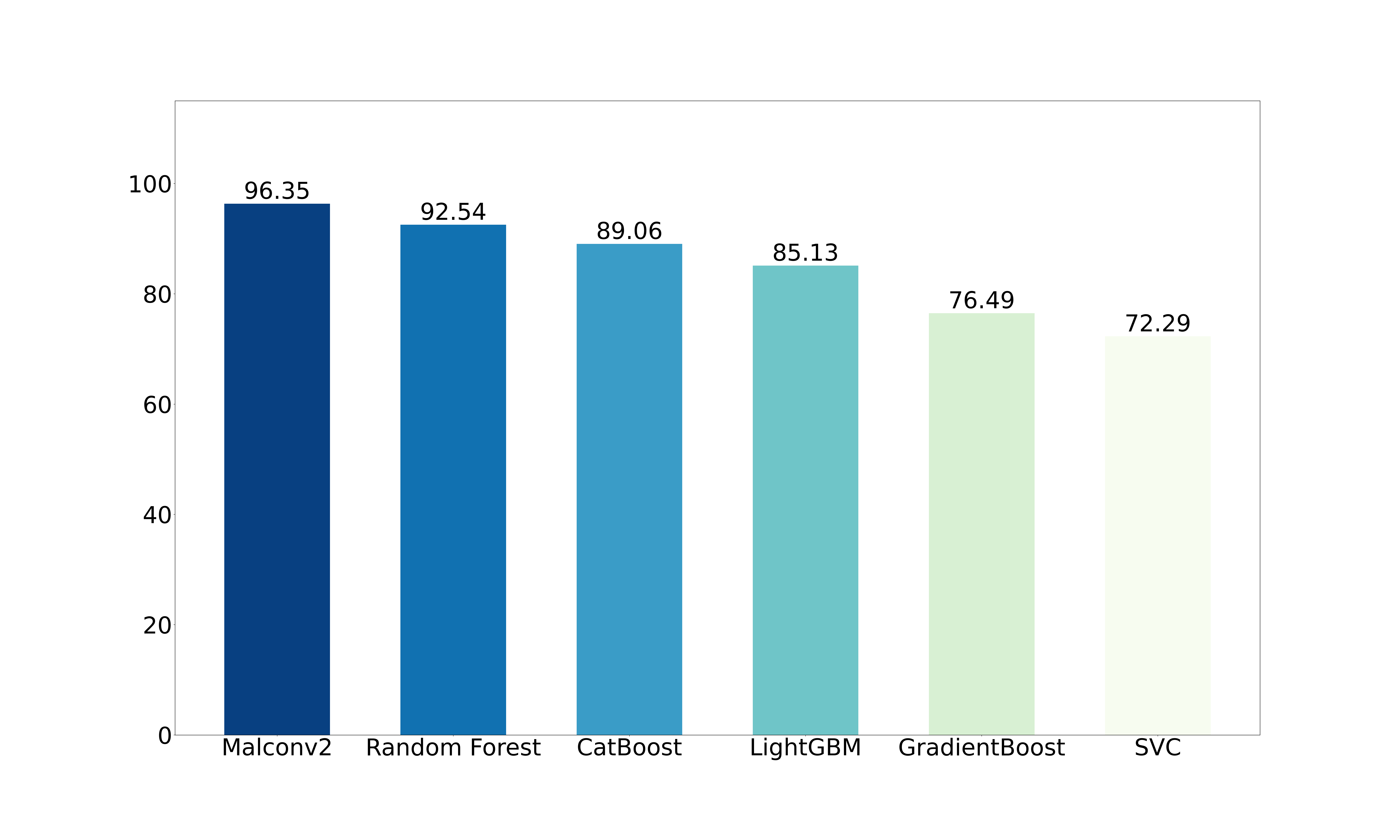}}
  \label{Fig:mlpre}
\end{figure}

\begin{figure}[H]
  \centering
  \captionof{figure}{ML/DL Model Recall on Test Data}\par\medskip
   \resizebox{\linewidth}{!}{
  \includegraphics[width=\linewidth]{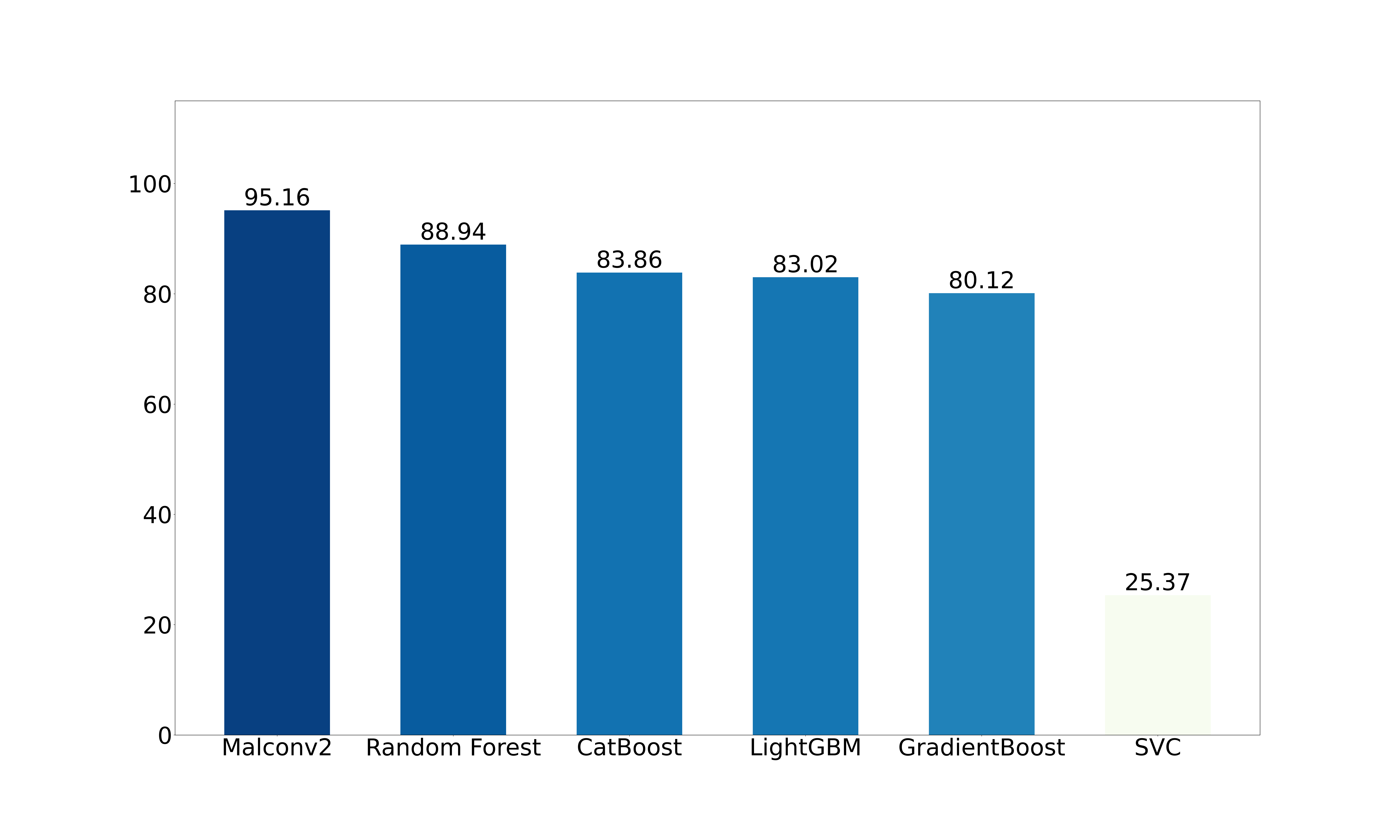}}
\label{Fig:mlrec}
\end{figure}

\begin{figure}[H]
  \centering
  \captionof{figure}{ML/DL Model F1 Score on Test Data}\par\medskip
   \resizebox{\linewidth}{!}{
  \includegraphics[width=\linewidth]{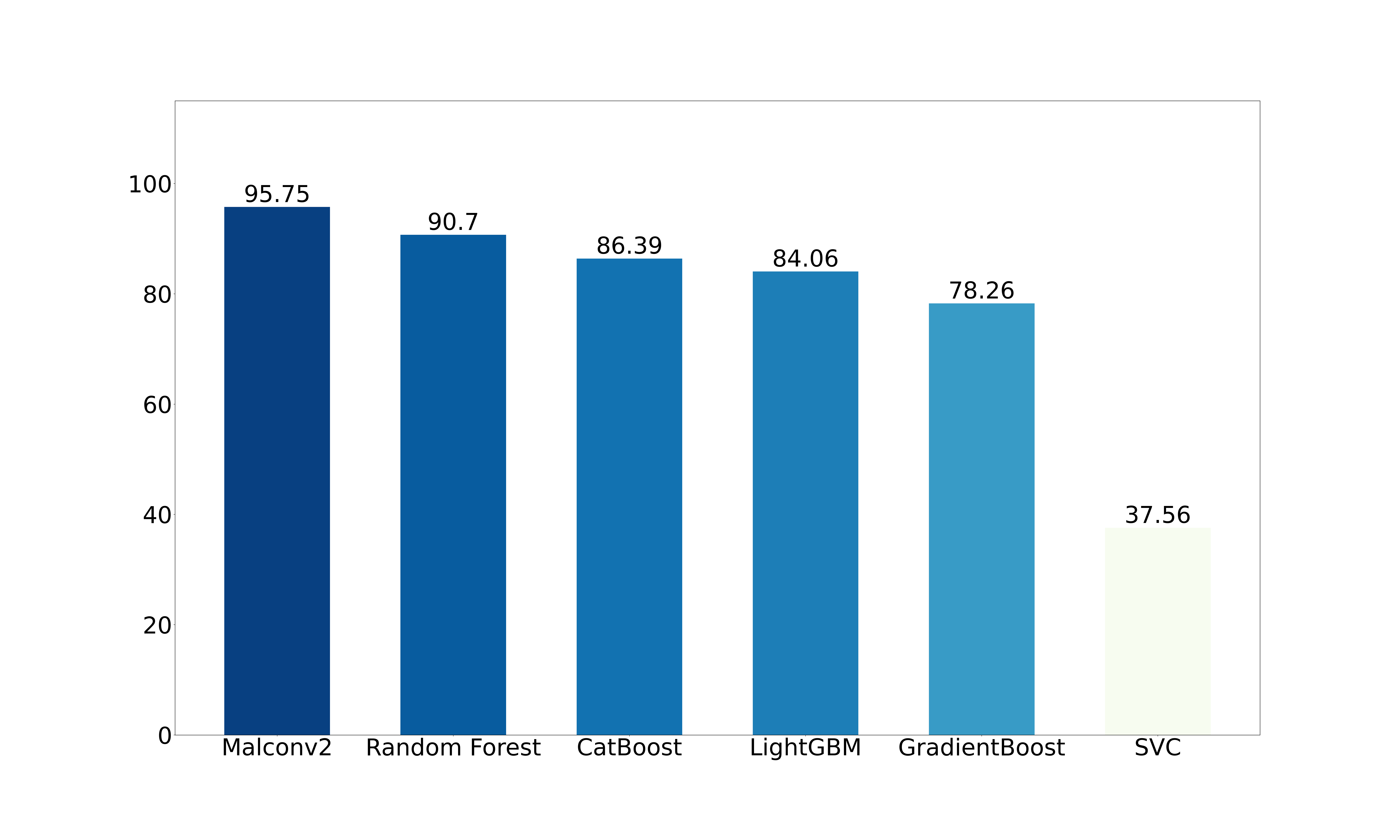}}
  \label{Fig:mlf1}
\end{figure}

Figures \ref{Fig:mlacc}, \ref{Fig:mlpre}, \ref{Fig:mlrec}, and \ref{Fig:mlf1} show the accuracy, precision, recall and F1 score, respectively, of a variety of machine learning models and one deep learning model (Malconv2). The ML models are trained on the byte histogram of each file while the deep learning model is trained on the raw bytes. 
%\bigskip
%\begin{minipage}{\linewidth}
%\centering
%\begin{table}[H]
%\captionof{table}{ML/DL Results on Test Data}
%\begin{tabular}{|l|l|l|l|l|}
%\hline
%Model                     & Accuracy & Precision & Recall         & F1 Score \\ \hline
%Malconv2                  & 96.23    & 96.35     & \textbf{95.16} & 95.75    \\ \hline
%Random Forest             & 91.85    & 92.54     & 88.94          & 90.70    \\ \hline
%CatBoost                  & 88.19    & 89.06     & 83.86          & 86.39    \\ \hline
%LightGBM                  & 85.93    & 85.13     & 83.02          & 84.06    \\ \hline
%GradientBoost             & 80.11    & 76.49     & 80.12          & 78.26    \\ \hline
%Support Vector Classifier & 62.30    & 72.29     & 25.37          & 37.56    \\ \hline
%\end{tabular}
%\label{Tab:ml}
%\end{table}
%\end{minipage}
%\bigskip

The deep learning model, Malconv2, had the highest recall and F1 score on our test data. However, the feature vector for the deep learning model is quite different from the feature vector for the ML models. Therefore, we attempted to improve the feature vector for our ML models by adding the mime-type output of \texttt{file} to the vector.

\begin{figure}[H]
  \centering
  \captionof{figure}{ML with \texttt{file} Accuracy on Test Data}\par\medskip
   \resizebox{\linewidth}{!}{
  \includegraphics[width=\linewidth]{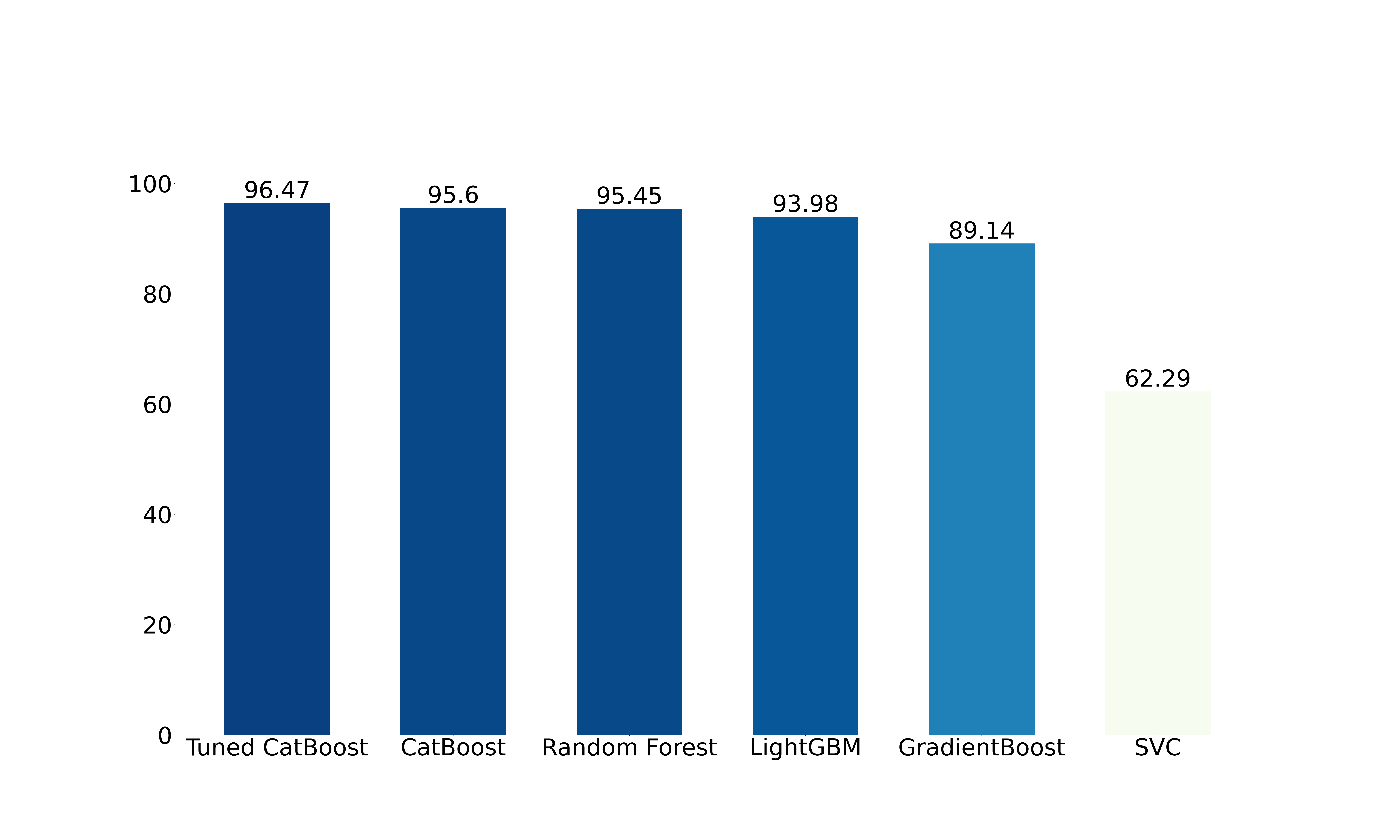}}
  \label{Fig:augacc}
\end{figure}

\begin{figure}[H]
  \centering
  \captionof{figure}{ML with \texttt{file} Precision on Test Data}\par\medskip
   \resizebox{\linewidth}{!}{
  \includegraphics[width=\linewidth]{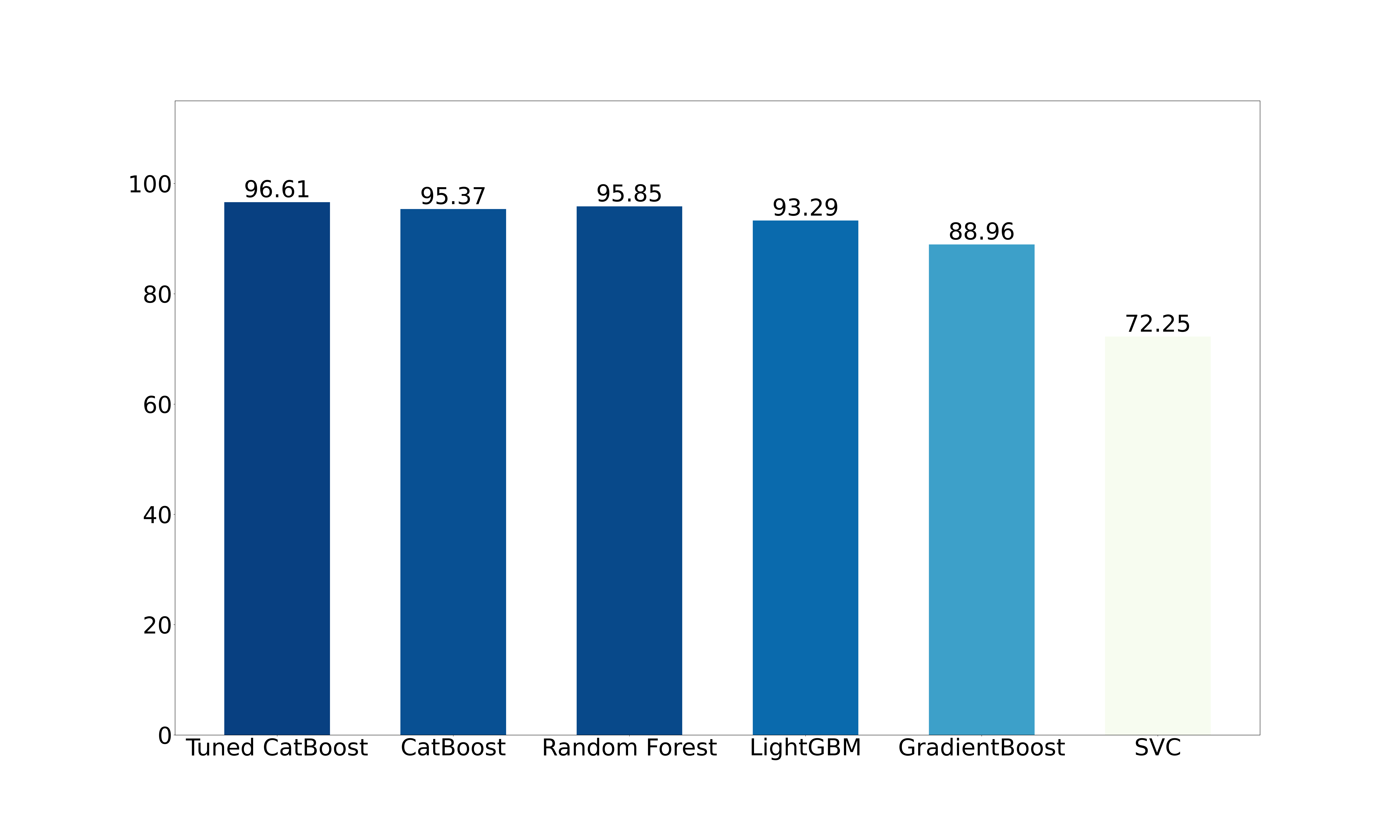}}
  \label{Fig:augpre}
\end{figure}

\begin{figure}[H]
  \centering
  \captionof{figure}{ML with \texttt{file} Recall on Test Data}\par\medskip
   \resizebox{\linewidth}{!}{
  \includegraphics[width=\linewidth]{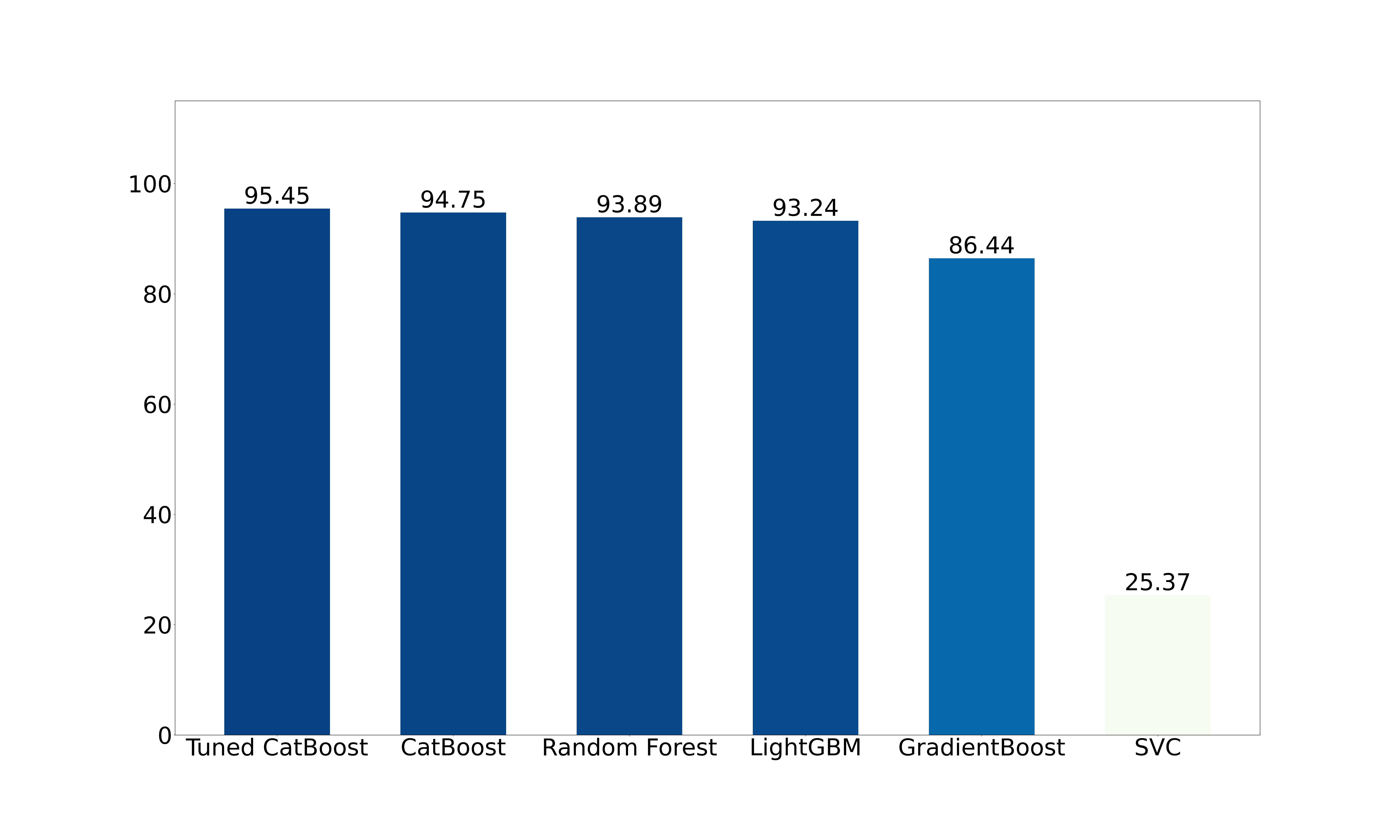}}
\label{Fig:augrec}
\end{figure}

\begin{figure}[H]
  \centering
  \captionof{figure}{ML with \texttt{file} F1 Score on Test Data}\par\medskip
   \resizebox{\linewidth}{!}{
  \includegraphics[width=\linewidth]{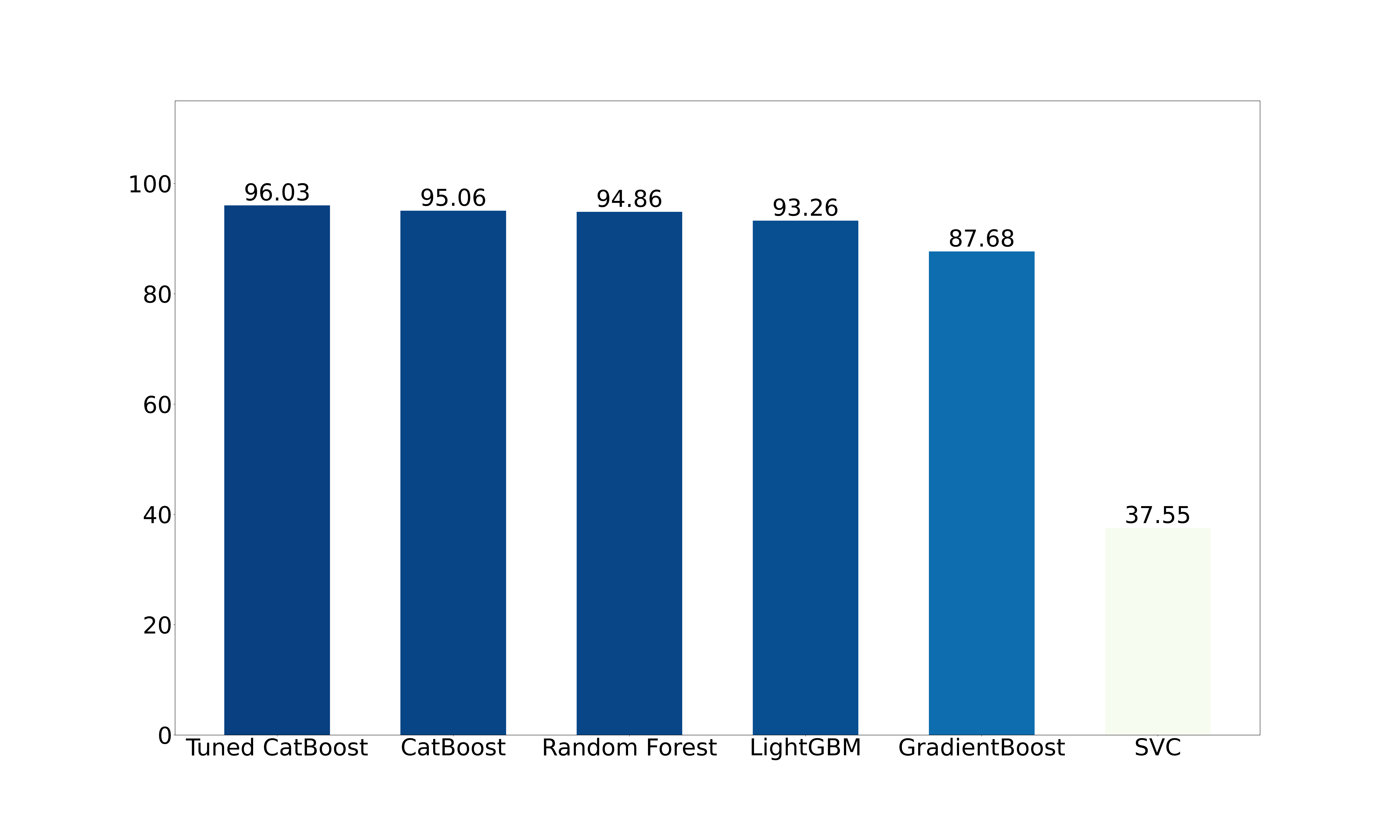}}
  \label{Fig:augf1}
\end{figure}

Tables \ref{Fig:augacc}, \ref{Fig:augpre}, \ref{Fig:augrec}, and \ref{Fig:augf1} show the accuracy, precision, recall, and F1 score of the ML models when the feature vector is a concatenation of the byte histogram and the mime-type (generated by \texttt{file}) for each file. The mime-type is converted to a 1-hot encoding for ingestion by the model. This additional feature boosted the recall of our machine learning models across the board. We tested whether adding this feature to Malconv2 would likewise boost the deep learning model's performance; however, the additional feature merely degraded Malconv2's performance. The performance of Catboost improved dramatically from 83.86 to 94.75. Due to this improvement, we tuned this model's hyperparameters to see if we could improve performance even more, yielding our best recall to date: 95.45.

The hyper-parameter tuned version of CatBoost narrowly outscored Malconv2 (95.45 vs 95.16 recall) thanks to the additional mime-type feature. The CatBoost model, like all of the ML models, has the advantage of a compact feature space compared to Malconv2, which expects raw bytes. Our version of Malconv2 truncates an input file if the file exceeds 16MB in size. Therefore, the feature vector is quite large (16MB) and training is much slower. Although much work remains to be done in terms of optimizing the detection of polyglot files, we currently recommend the use of a Catboost model whose feature space consists of the byte histogram concatenated with the mime-type output of file (converted to a 1-hot encoding) for polyglot detection. This model had the highest recall combined with an efficient feature space.

%\bigskip
%\begin{minipage}{\linewidth}
%\begin{table}[H]
%\centering
%\captionof{table}{ML with \texttt{file} Results on Test Data}
%\begin{tabular}{|l|l|l|l|l|l}
%\cline{1-5}
%Model                     & Accuracy & Precision & Recall         & F1 Score &  \\ \cline{1-5}
%Tuned CatBoost            & 96.47    & 96.61     & \textbf{95.45} & 96.03    &  \\ \cline{1-5}
%CatBoost                  & 95.60    & 95.37     & 94.75          & 95.06    &  \\ \cline{1-5}
%Random Forest             & 95.45    & 95.85     & 93.89          & 94.86    &  \\ \cline{1-5}
%LightGBM                  & 93.98    & 93.29     & 93.24          & 93.26    &  \\ \cline{1-5}
%GradientBoost             & 89.14    & 88.96     & 86.44          & 87.68    &  \\ \cline{1-5}
%Support Vector Classifier & 62.29    & 72.25     & 25.37          & 37.55    &  \\ \cline{1-5}
%\end{tabular}
%\label{Tab:ml_file}
%\end{table}
%\end{minipage}
%\bigskip

\section{Conclusion \& Future Work}
We demonstrated that the most common method for file type identification fails to reliably detect polyglots. We then evaluated machine and deep learning models for detecting polyglots and found that Malconv2 had the highest recall. If mime-type is included as a feature, then a hyper-parameter tuned version of CatBoost has the highest recall. CatBoost has the added advantage of a much smaller feature vector. Our work is merely the beginning, however. There are several avenues of research that need to advanced in order to produce an operational polyglot detector, namely:
\begin{itemize}
    \item Data diversification: A production-grade system can encounter an extremely wide variety of file formats; ergo, We need to assemble a data set that fully encompasses the wide variety of file formats commonly encountered in the wild. 
    \item Feature space optimization: We need to asses and possibly incorporate additional features into our feature vector. Historical candidates include byte co-occurrence counts and byte entropy.
    \item Hyper-parameter tuning: We need to continue to optimize the top-scoring models to achieve maximum performance.
    \item High throughput: We need to demonstrate that our top-scoring models have high enough throughput to be incorporated into production-grade systems where high throughput is essential.
    \item Benign polyglots: We need to address instances where files are cmobined into polyglots for benign purposes, e.g., DICOM+TIFF files that combine medical images into one file.
\end{itemize}

%We need to demonstrate that the above models have high enough throughput to be incorporated into a production-grade malware detection system. We are in the process of conducting feature space optimization to determine if other features will enhance performance. We also plan to conduct extensive hyper-parameter tuning to improve model performance. Moreover, we plan to generate a larger data set with a much wider variety of donor types to establish whether our approach is valid for a detection system than can encounter a practical variety of file types. Finally, we need to address the instances where polyglots are deployed in a non-malicious manner, as in a DICOM+TIFF.

%\section*{Artifact Availability}
%The byte count and mime type data for both monoglot and polyglot files used to train the machine learning models is available at [redacted].

\bibliographystyle{IEEEtran}
\bibliography{references.bib}

\end{document}